\documentclass[11pt]{article}

\usepackage{mathbbol}
\usepackage{caption}
\usepackage{graphicx}
\usepackage{ushort}
\usepackage{amsmath}
\usepackage{amssymb}
\usepackage{makeidx}
\usepackage{float}
\usepackage{accents}
\usepackage{framed}
\usepackage{versions}
\usepackage{emptypage}
\usepackage{graphicx}
\usepackage{amsfonts,amsthm}

\def\hmath$#1${\texorpdfstring{{\rmfamily\textit{#1}}}{#1}}

\usepackage[usenames,dvipsnames]{color}
\usepackage{hyperref}

\newcommand{\RL}{{\mathbb R}}

\newcommand{\IN}{{\mathbb Z}}

\newcommand{\IND}{{\mathbb I}}

\newcommand{\PB}{\mbox{\boldmath $P$}}

\def\ba{\begin{align}}
\def\ea{\end{align}}
\def\ban{\begin{align*}}
\def\ean{\end{align*}}

\def\be{\begin{eqnarray}}
\def\ee{\end{eqnarray}}
\def\ben{\begin{eqnarray*}}
\def\een{\end{eqnarray*}}

\def\bqq{\begin{equation}}
\def\eqq{\end{equation}}
\def\bqqn{\begin{equation*}}
\def\eqqn{\end{equation*}}

\def\elabel#1{\label{e:#1}}

\def\sq{$\Box$}

\def\qed{\ifmmode\sq\else{\unskip\nobreak\hfil
\penalty50\hskip1em\null\nobreak\hfil\sq
\parfillskip=0pt\finalhyphendemerits=0\endgraf}\fi\par\medbreak}

\newsavebox{\junk}
\savebox{\junk}[1.6mm]{\hbox{$|\!|\!|$}}

\def\bfW{{\bf W}}

\def\bfPhi{\mbox{\protect\boldmath$\Phi$}}
\def\bfPsi{\mbox{\protect\boldmath$\Psi$}}

\def\til={{\widetilde =}}

\def\clD{{\cal D}}

\def\clL{{\cal L}}

 \def\eq#1/{(\ref{#1})}

\theoremstyle{plain}

\newtheorem{theorem}{Theorem}[section]
\newtheorem{corollary}[theorem]{Corollary}
\newtheorem{proposition}[theorem]{Proposition}
\newtheorem{lemma}[theorem]{Lemma}

\def\eq#1/{(\ref{e:#1})}

\newcommand{\beqn}[1]{\notes{#1}%
\begin{eqnarray} \elabel{#1}}

\newcommand{\eeqn}{\end{eqnarray} } 

\newcommand{\beq}[1]{\notes{#1}%
\begin{equation}\elabel{#1}}

\newcommand{\eeq}{\end{equation}} 

\def\bdes{\begin{description}}
\def\edes{\end{description}}

\def\notes#1{}

\definecolor{mag}{rgb}{0.7,0,0.3}
\definecolor{dgreen}{rgb}{0.1,0.5,0.1}
\definecolor{dred}{rgb}{.8,0,0}
\definecolor{gray}{rgb}{.8,.8,.8}
\definecolor{brown}{rgb}{0.6451,0.3706,0.1745}

\def\bfUpsilon{\mbox{\protect\boldmath$\Upsilon$}}
\def\bftPsi{\mbox{\protect\boldmath$\tilde{\Psi}$}}

\begin{document}
 
\title{\vspace{-2cm}%
A Simple Network of Nodes Moving on the Circle
}
\author{
	D. Cheliotis\thanks{Department of Mathematics,
	Panepistimiopolis, Athens 15784, Greece.
        Email: {\tt dcheliotis@math.uoa.gr}.
	}
\and
      	I. Kontoyiannis\thanks{Department of Engineering,
		University of Cambridge,
		Trumpington Street, 
		Cambridge CB2 1PZ, UK;
		and
		Department of Informatics,
		Athens University of Economics and Business,
		Patission 76, Athens 10434, Greece.
                Email: {\tt ik355@cam.ac.uk}.
	}
\and
	M. Loulakis\thanks{
 		School of Applied Mathematical and Physical Sciences, 
		National Technical University of Athens, 
		Polytechnioulopis Zographou, Athens 15780, Greece;
		and Institute of Applied and Computational Mathematics, 
		Foundation for Research and Technology - Hellas, Greece.
		Email: {\tt loulakis@math.ntua.gr}.
	}
\and
	S. Toumpis\thanks{
		Department of Informatics,
		Athens University of Economics and Business,
		Patission 76, Athens 10434, Greece.
                Email: {\tt toumpis@aueb.gr}.
	}
}

\maketitle
\thispagestyle{empty}
\setcounter{page}{0}

\footnotetext{Preliminary versions of some of the present results
appeared in the conference papers \cite{CKLT:17,CKLT:18}.}

\footnotetext{This research work was supported by the Hellenic Foundation for 
Research and Innovation (H.F.R.I.) under the ``First Call for H.F.R.I.\
Research Projects to support Faculty members and Researchers and the 
procurement of high-cost research equipment grant,'' project number 1034.}
 
\begin{abstract}
Two simple Markov processes are examined, one in discrete and one in continuous
time, arising from idealized versions of a transmission protocol for mobile
networks. We consider two independent walkers moving with constant speed on the
discrete or continuous circle, and changing directions at independent 
geometric (respectively, exponential) times. One of the walkers carries 
a message that wishes to travel as
far and as fast as possible in the clockwise direction. The message 
stays with its current
carrier unless the two walkers meet, the carrier is moving 
counter-clockwise, and the other
walker is moving clockwise. Then the message jumps to the other walker. 
Explicit expressions are
derived for the long-term average clockwise speed and number of jumps made 
of the message,
via the solution of associated boundary value problems.
The tradeoff between speed and cost (measured as the rate of jumps) 
is also examined.

\bigskip


\bigskip


\end{abstract}

\thispagestyle{empty}

\newpage

\section{Introduction}
\label{s:intro}

Consider a network that consists of many mobile nodes moving 
around randomly in some large area.
Suppose that each node moves with constant speed, 
changing its direction of travel at random times,
and that one of the nodes carries a message that 
she wants to transmit to a far away destination in some specific,
fixed
direction. The message stays with its current carrier until the 
first time she comes within a certain distance from some other
node moving in a `better' direction, i.e., in a direction 
closer to that of the intended recipient.
In that case,
she transmits her message to the other node, and the new carrier
then proceeds in the same fashion. What is the long-term average 
speed with which the message travels towards its destination, 
as a function of, say, the nodes' individual speeds and their 
density? How often, on the average, does the message get 
transmitted form one node to another?

Networks of this type, where messages propagate via
a combination of physical transport (moving with their carrier)
and wireless transmissions (being sent from one node to another)
belong to the wide class of delay-tolerant networks 
(DTNs)~\cite{dtn_book}. Examples of DTNs arising in applications
include
space~\cite{araniti2015}, vehicular~\cite{baccelli78},
sensor~\cite{pathirana2005}, and pocket-switched networks~\cite{hui2011}. 
In earlier work by some of the authors
\cite{cavallari1,cavallari2,cavallari3},
the questions of the previous paragraph
were considered 
under very general assumptions on the movement
of the nodes and on the protocol under which 
the message gets transmitted between nodes. In that line of work,
as in much of the related earlier work in this area,
e.g., \cite{tse1,diggavi1,jacquet17,peres1},
the complexity 
of the models involved prohibits the derivation of exact, 
explicit answers. For that reason, typical results
are in the form of asymptotics, approximations,
performance bounds, or estimates based on
simulation experiments.

In this work we examine two variants of a simple, idealized model, 
where it is possible to derive explicit, closed-form 
expressions for the performance metrics of interest. 
We first consider a collection of $m\geq 2$ nodes moving
independently on a discrete circle consisting of $N\geq 3$
locations, in discrete time. Each node maintains their current direction
of travel for a geometrically distributed amount of time
with parameter $\epsilon\in(0,1)$, and one of the nodes
carries a message intended to travel as far as possible
in the clockwise direction. The message stays with its
current carrier unless, while moving
counter-clockwise, it finds itself in the same location
as a different node moving clockwise. In that case
the message gets transmitted to the other node,
and the same process is repeated.

For the case of $m=2$ nodes, in 
Section~\ref{s:disc_res},
Theorem~\ref{thm:speed2},
we show that the long-term average clockwise speed 
$s=s(N,\epsilon)$ 
of the message is
$s =\frac{1-\epsilon}{2(1+\epsilon(N-2))}.$
The proof, given in Section~\ref{s:proofs},
involves the construction of a martingale
that solves an associated (discrete) boundary value problem.
Similar techniques allow us to compute the average
transmission cost $c=c(N,\epsilon)$,
measured as the long-term average number of message 
transmissions per unit time. 
In Theorem~\ref{thm:cost2} we show that
$c(N,\epsilon)=\epsilon \times s(N,\epsilon)$,
for all $N$ and~$\epsilon$. Therefore, the message travels
a clockwise distance of $1/\epsilon$ units
between successive jumps (on the average),
regardless of $N$. The tradeoff between
speed and cost for different values of the parameters
$N$ and $\epsilon$ is also discussed in Section~\ref{s:disc_res}.

Section~\ref{s:cont_res} contains continuous-time
analogs of Theorems~\ref{thm:speed2} and~\ref{thm:cost2}.
Here we consider $m=2$ nodes moving with constant speed $v>0$
on a continuous circle of circumference $N>0$, changing directions 
at independent exponential times with rate $r>0$. 
The corresponding expressions for the long-term
average speed $s(N,v,r)$
and cost $c(N,v,r)$ are established in Theorems~\ref{thm:Cspeed}
and~\ref{thm:Ccost2}, respectively. Again, it turns out that
the speed and cost satisfy a simple scale-free relationship:
$c(N,v,r) = (r/v)\times s(N,v,r)$. In other words, the message
travels a (clockwise) distance of $v/r$ units between successive jumps,
on the average.

The proofs of
Theorems~\ref{thm:Cspeed} and~\ref{thm:Ccost2},
given in Section~\ref{s:proofs}, involve shorter
and somewhat cleaner arguments than their discrete-time 
counterparts. In the continuous-time case, it is more
straightforward to construct appropriate
solutions to the relevant boundary value problems,
which are stated in terms of the infinitesimal 
generator of the underlying Markov process.
What is somewhat cumbersome, is the proof 
that this Markov process is exponentially ergodic, 
uniformly in its initial state. The relevant 
ergodic properties are stated in Proposition~\ref{prop:ergodic0} and 
Theorem~\ref{thm:ergodic}, both proved in the Appendix.

Although perhaps the most restrictive of our 
assumptions is that nodes are assumed to move 
along the circumference of a circle, we note that
there has been much recent interest in one-dimensional
models of DTNs, particularly in connection with 
the important class of vehicular networks (VANETS);
see \cite{baccelli78,baccelli79,zarei17} and the references
therein. Finally, a somewhat less closely related
but quite extensively studied problem, in terms of 
a Markov chain describing the movement of a finite
collection of nodes on a circle, is the 
{\em $k$-server} problem
introduced in \cite{manasse:88};
see, e.g.,  \cite{csaba:06} or \cite{bansal:15}
for more recent developments.

\section{Models and Problem Statement}
\label{s:models}

\subsection{Random walk on the discrete circle}
\label{s:problemD}

Let $S:=\{0,1,\ldots,N-1\}=\IN/N\IN$ denote the discrete
$N$-circle, for a fixed odd $N\geq 3$. We place
$m\geq 2$ independent
random walkers on $S$, located 
at $X_t=(X_t(1),X_t(2),\ldots,X_t(m))$ at time $t=0,1,2,\ldots$,
and with each walker $j$ we associate a random direction
$D_t(j)$ at time $t$, where $D_t(j)$ is either $=+1$
(clockwise motion) or $D_t(j)=-1$ (counter-clockwise motion).
The initial positions $X_0$ and directions $D_0$ are
arbitrary.
The Markov chain $\{(X_t,D_t)\;;\;t\geq 0\}$
evolves on the state space
$S^m\times\{-1,+1\}^m$ as follows.

Let 
$\{Z_t=(Z_t(1),Z_2(2),\ldots,Z_t(m))\}$
be a sequence of
independent Bernoulli random variables 
with parameter $\epsilon\in(0,1)$.
Given the current state
$(X_t,D_t)$, each walker $j$ takes a
step in the direction given by $D_t(j)$,
$$X_{t+1}(j)=X_t(j)+D_t(j)\pmod N,\;\;\;t\geq 0,\;j=1,2,\ldots,m,$$
and then decides to either continue moving in the
same direction with probability $(1-\epsilon)$,
or to switch to the opposite direction, with probability
$\epsilon$:
$$D_{t+1}(j)=(1-Z_t(j))D_t(j)-Z_t(j)D_t(j),\;\;\;t\geq 0,\;j=1,2,\ldots,m.$$

We also define an index process
$\{I_t\}$ evolving on $\{1,2,\ldots,m\}$,
with $I_0$ chosen arbitrarily
and $I_t$ trying to track walkers that move
clockwise: Given $(X_t,D_t,I_t=i)$, let
$(X_{t+1},D_{t+1})$ be defined as above.
If $D_{t+1}(i)=-1$ and there
is at least one more walker, $j$, say, 
at the same location, $X_{t+1}(i)=X_{t+1}(j)$,
but its direction $D_{t+1}(j)=+1$, then 
$I_{t+1}=j$ (or a uniformly chosen
such $j$ if there are multiple candidates). 
In all other cases, $I_{t+1}=I_t=i$.

It is easy to see from the above construction
that $\bfPhi=\{\Phi_t=(X_t,D_t,I_t)\;;\;t\geq 0\}$ is an
irreducible and aperiodic chain on the
state space $\Sigma$ consisting of 
all configurations of the form,
$$(x(1),x(2),\ldots,x(m),d(1),d(2),\ldots,d(m),i)\in
S^m\times\{+1,-1\}^m\times\{1,2,\ldots,m\},$$
except those where $d(i)=-1$ and there is a $j\neq i$
such that $x(j)=x(i)$ and $d(j)=+1$.
Moreover, under the unique invariant distribution $\pi$
of $\bfPhi$, the distribution of 
$(X_t,D_t)$ is uniform: 
The positions $X_t(i)$ are
independent of each other and uniformly distributed on $S$,
and the directions $D_t(i)$
are independent
of the positions $X_t$ and each 
$D_0(i)=\pm1$ with probability $1/2$,
independently of the others. 

We are primarily interested in the following 
three quantities, as functions of $N,m$ and~$\epsilon$:
$(i)$~{\em Direction}: What is the limiting distribution 
of the 
direction $D_t(I_t)$ of the message at time~$t$?
$(ii)$~{\em Speed}: What is
the long-term average speed of
the message? 
$(iii)$~{\em Cost}:
What is the long-term average number
of jumps per unit time?
We are also interested in the
relationship between the speed and cost:
Do higher speeds always imply an increase in cost?
Or is there a range of parameter values that
improve the speed and cost simultaneously?

\subsection{Continuous motion on the circle}
\label{s:problemC}

Let $S:=\RL/N\IN$ denote the one-dimensional circle
of circumference $N>0$, where $N$ is not necessarily
an integer.
We place $m\geq 2$ independent
random walkers $X_t=(X_t(1),\ldots,X_t(m))$ on $S$,
and with each walker $j$ we associate a random direction
$D_t(j)$ at time $t$, where $D_t(j)$ is either $=+1$
(clockwise motion) or $D_t(j)=-1$ (counter-clockwise motion).
The initial positions $X_0$ and directions $D_0$ are
arbitrary.
The continuous-time Markov process $\bfW=\{(X_t,D_t)\;;\;t\geq 0\}$
evolves on the state space
$S^m\times\{-1,+1\}^m$ as follows.
The $j$th walker continues moving at constant speed $v$
in its present direction, $D_t(j)=d$, say, for an exponentially 
distributed amount of time with mean $1/r$,
for some $r>0$; during that time
its direction remains constant,
and afterwards it switches to $-d$.
The
process continues in the same fashion,
by choosing a new, independent exponential 
time for the $j$th walker, and with the
different walkers moving independently of
one another.

We assume that the transitions between
directions are such that the sample paths
of 
the process $\bfW=\{(X_t,D_t)\;;\;t\geq 0\}$ 
are right
continuous, and observe that $\bfW$ is 
strong Markov and, therefore, a Borel
right process \cite{sharpe:book,dowmeytwe}.
And since 
$S^m\times\{-1,+1\}^m$ 
is compact, $\bfW$ is also non-explosive
\cite{meytwe93b}. The following simple
proposition is proved in the Appendix.

\begin{proposition}
\label{prop:ergodic0}
\begin{itemize}
\item[$(i)$] The Markov process $\bfW=\{(X_t,D_t)\;;\;t\geq 0\}$
is $\psi$-irreducible and aperiodic 
on $S^m\times\{-1,+1\}^m$, with respect to
$\psi:=\clL^m\times\kappa^m$, where $\clL$
denotes the Lebesgue measure on $S$ 
and $\kappa$ the counting measure on $\{+1,-1\}$.
\item[$(ii)$] 
The process $\bfW$ is positive Harris recurrent.
\item[$(iii)$] 
The uniform distribution is 
the unique invariant probability measure of $\bfW$.
\end{itemize}
\end{proposition}

We also define an index process
$\{I_t\}$ evolving on $\{1,2,\ldots,m\}$,
with $I_0$ chosen arbitrarily
and $I_t$ trying to track walkers that move
clockwise. Specifically, $I_t$ stays constant
most of the time, and its value only changes
when $X_t(I_{t-})=X_t(j)$ for some $j\neq I_t$, and the direction $D_t(j)$ 
of the $j$th walker at the time is +1 while $D_t(I_{t-})=-1$. 
In that
case, the value of $I_t$ switches to $j$
(or to a uniformly chosen such $j$ if there 
are multiple candidates) and
remains there at least until the first time
walker $j$ encounters a different walker.

Next we show that
the Markov process
$\bfPhi=\{\Phi_t=(X_t,D_t,I_t)\;;\;t\geq 0\}$ is 
uniformly ergodic
on the state space $\Sigma$, which 
consists of all elements 
$\phi\in S^m\times\{-1,+1\}^m\times\{1,2,\dots,m\}$,
$$\phi=(x(1),x(2),\ldots,x(m),d(1),d(2),\ldots,d(m),i),$$
where we identify pairs of states $s=(x,d,i)$
and $s'=(x',d',i')$ of the following form:
The message is with a different walker 
in each state, i.e., $i\neq i'$,
all positions and directions are identical,
$x=x'$ and $d=d'$, the 
$i$th and $i'$th walkers are in the same position
$x(i)=x(i')$, 
and the two walkers move in opposite directions,
i.e.,
$d(i)=-d(i')$.

As with $\bfW$, we assume that the transitions 
between directions and between successive
values of the process $\{I_t\}$
are such that the sample paths 
of $\bfPhi$ are right-continuous,
so that 
$\bfPhi$ is a
non-explosive, Borel
right process \cite{sharpe:book,dowmeytwe}.
Its ergodicity properties are 
summarized in Theorem~\ref{thm:ergodic}, proved 
in the Appendix.
Here,
and throughout the paper, for an arbitrary measure $\mu$
and function $g$ we write $\mu(g)$ for $\int gd\mu$, whenever
the integral exists.

\begin{theorem}
\label{thm:ergodic}
\begin{itemize}
\item[$(i)$]
$\bfPhi=\{\Phi_t=(X_t,D_t,I_t)\;;\;t\geq 0\}$ is 
$\psi$-irreducible and aperiodic with respect to
$\psi:=\clL^m\times\kappa^m\times\kappa_m$, where,
as before, $\clL$ and $\kappa$ denote the Lebesgue 
and counting measures on $S$ and $\{+1,-1\}$, respectively,
and $\kappa_m$ denotes the counting measure on $\{1,2,\ldots,m\}$.
\item[$(ii)$]
$\bfPhi$ is
uniformly ergodic, with a unique invariant
probability measure $\pi$.
\item[$(iii)$]
$\bfPhi$ converges to equilibrium uniformly exponentially
fast: There are constants $C<\infty,\rho>0$ such that,
$$|P_\phi(\Phi_t\in A)-\pi(A)|=|\Pr(\Phi_t\in A|\Phi_0=\phi)-\pi(A)|\leq Ce^{-\rho t},$$
for all $\phi\in\Sigma$, all measurable $A\subset\Sigma$, 
and all $t>0$.
\item[$(iv)$]
The following ergodic theorem holds for $\bfPhi$: 
For any bounded (measurable) function $f:\Sigma\to\RL$
and any initial state $\Phi_0=\phi\in\Sigma$,
$$\lim_{t\to\infty}\frac{1}{t}\int_0^t f(\Phi_s)ds=\pi(f),\;\;\;a.s.$$
\end{itemize}
\end{theorem}

Finally we note that the dynamics 
of $\bfPhi$ can be described by its 
infinitesimal generator $L$.
Although we will not give a complete
description of $L$, we note the following
fact which is easy to establish and which
we will need later.
For any function
$f:\Sigma\to\RL$ in the domain of $L$,
which is continuously differentiable in $x$,
the value of $Lf$ at a pont $(x,d,i)$
with $x(j)\neq x(i)$ for all $j\neq i$ is, 
\be
Lf(x,d,i)=\sum_{j=1}^m 
\left\{v d(j)\frac{\partial f}{\partial x(j)}(x,d,i)+r\left[
f(x,\sigma^jd,i)-f(x,d,i)
\right]
\right\},
\label{eq:generator}
\ee
where, for any $m$-tuple of directions $d\in\{-1,+1\}^m$,
$\sigma^jd$ is the same as $d$ but with
its $j$th coordinate having the opposite sign
from that of $d$,
$1\leq j\leq m$.
The first term in the sum on the right-hand side above
corresponds to the motion of the $j$th walker at constant 
velocity $vd(j)$, while the second one corresponds to 
its change of direction at rate $r$.

Once again,
we are interested in the following three quantities,
as functions of~$m,N,v$ and~$r$:
$(i)$~The limiting distribution 
of the 
direction $D_t(I_t)$ of the message at time~$t$;
$(ii)$~The long-term average speed of
the message;
$(iii)$~The long-term average number
of jumps per unit time.
Also, we wish to examine the nature of
the tradeoff between the speed and cost.


\section{Results: Speed and Cost with \hmath $m=2$ walkers}
\label{s:results}

Here we state and discuss our main results for both the discrete
and the continuous case. The proofs are given in Section~\ref{s:proofs}.
We adopt the following standard notation: For the probabilities of
events depending on an underlying Markov process $\{\Phi_t\}$ we write
$P_\phi$ for the measure describing the distribution of the process
conditional on $\{\Phi_0=\phi\}$, and $P_\mu$ when $\Phi_0\sim\mu$
for some probability measure $\mu$. Similarly, $E_\phi$ and $E_\mu$
denote the corresponding expectation operators.

\subsection{The discrete circle}
\label{s:disc_res}

Consider the problem of $m=2$ walkers on the $N$-circle,
changing directions with rate $\epsilon$, as described
in Section~\ref{s:problemD}.

\begin{theorem}[{\bf Message speed}]
\label{thm:speed2}
In the case of $m=2$ walkers, for any initial state,
the long-term average speed of the message is:
$$s:=\pi(D_1(I_1))=\lim_{n\to\infty}\frac{1}{n}\sum_{t=0}^{n-1}D_t(I_t)
=\frac{1-\epsilon}{2(1+\epsilon(N-2))},
\;\;\;\mbox{a.s.}$$
\end{theorem}

\begin{figure}[ht!]
\centerline{\includegraphics[width=6.1in]{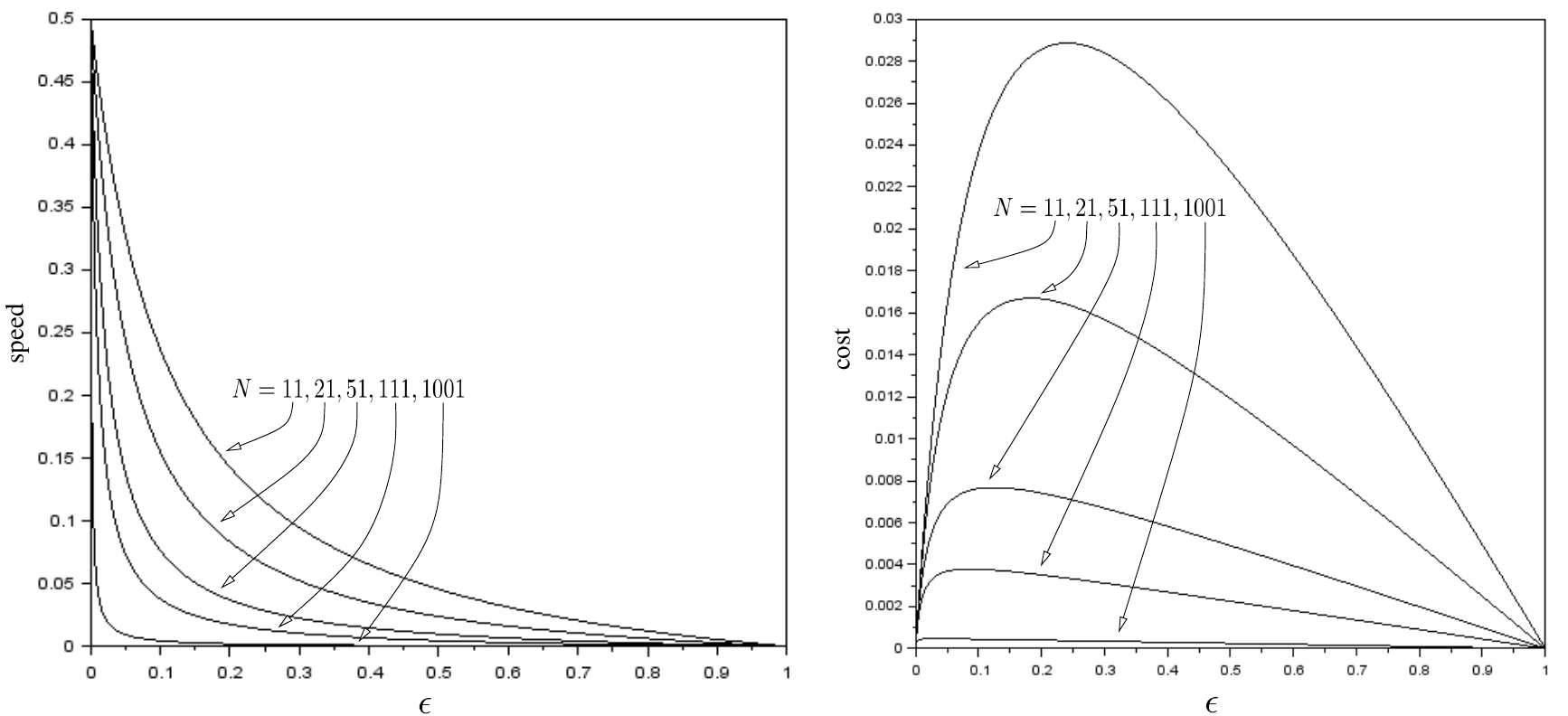}}
\caption{Plots of the asymptotic speed $s=s(N,\epsilon)$
of the message (left) and 
of the asymptotic cost $c=c(N,\epsilon)$ (right),
as functions of $\epsilon$, for different values of $N$.}
\label{f:theory2}
\end{figure}

\noindent
Note that the speed $s=s(N,\epsilon)$ is 
always less than or equal to $1/2$,
and it is decreasing in both $N$ and $\epsilon$;
see Figure~\ref{f:theory2}. 

\begin{figure}[ht!]
\centerline{\includegraphics[width=6.3in]{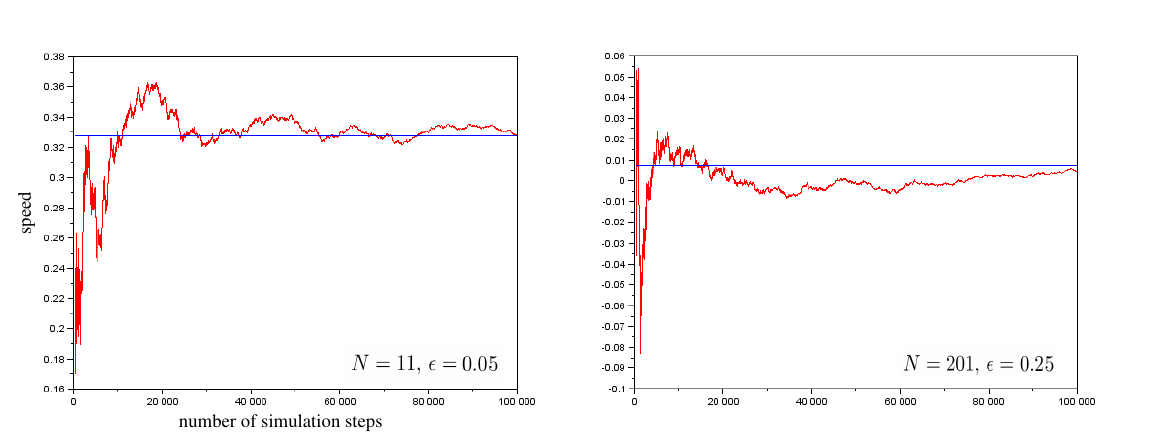}}
\caption{Simulation results for the speed of the message 
during $T=10^5$ steps, in two independent 
realizations of the chain with different
parameter values. In each case, the horizontal line 
is the limiting value of the speed $s$ predicted
by Theorem~\ref{thm:speed2}.}
\label{f:speed2}
\end{figure}

In the boundary case
$\epsilon=0$, the speed $s(N,\epsilon)$  is either -1 or 1, 
depending on the initial directions of the two walkers.
Therefore, $s(N,\epsilon)$ is discontinuous at $\epsilon=0$,
since $s(N,\epsilon)\uparrow 1/2$ as $\epsilon\downarrow 0$,
for any $N$.
Figure~\ref{f:speed2} shows the results of
two simulation experiments, illustrating the
convergence of the speed of the message to the
corresponding value $s(N,\epsilon)$
computed in Theorem~\ref{thm:speed2}. 

Theorem~\ref{thm:speed2} answers
question~$(ii)$ of 
Section~\ref{s:problemD}.
The answer to question~$(i)$ 
is a simple consequence of the theorem,
given in Corollary~\ref{cor:direction2}
below.

\begin{corollary}[{\bf Message direction}]
\label{cor:direction2}
In the case of $m=2$ walkers, for any initial state $\Phi_0=\phi\in\Sigma$,
the steady state probability that the message  moves in 
the clockwise direction is:
$$P_\pi(D_1(I_1)=+1)=\lim_{t\to\infty}P_\phi(D_t(I_t)=+1)
=\frac{s+1}{2}=\frac{3+\epsilon(2N-5)}{4(1+\epsilon(N-2))}.$$
\end{corollary}

\noindent
Next we examine the asymptotic cost of 
message transmissions. Theorem~\ref{thm:cost2} describes
the long-term average number of jumps of the message,
$c=c(N,\epsilon)$ per unit time.

\begin{theorem}[{\bf Transmission cost}]
\label{thm:cost2}
In the case of $m=2$ walkers, for any initial state,
the long-term average cost of message transmissions is:
$$c:=P_\pi(I_2\neq I_1)=\lim_{n\to\infty}
\frac{1}{n}\sum_{t=0}^{n-1}
\IND_{\{I_{t+1}\neq I_t\}}
=\frac{\epsilon(1-\epsilon)}{2[1+\epsilon(N-2)]},
\;\;\;\mbox{a.s.}$$
\end{theorem}

We observe that the cost $c=c(N,\epsilon)$ is decreasing in $N$,
and for each fixed $N$ it is a concave function of~$\epsilon$; 
see Figure~\ref{f:theory2}.
Also, unlike the speed $s=s(N,\epsilon)$, the cost
$c=c(N,\epsilon)$ is continuous and
equal to zero at $\epsilon=0$.

\medskip

\noindent
{\bf Speed vs.\ cost. }
It is interesting to observe the
following simple, scale-free relationship
between the asymptotic speed and 
cost: $c(N,\epsilon)=\epsilon \times s(N,\epsilon)$,
for all $N$ and~$\epsilon$.
Therefore, on the average, the message travels a 
(clockwise) distance of $1/\epsilon$ units
between successive jumps, regardless of the
value of $N$.

In terms of the speed/cost tradeoff,
note that for each $N$ there is an $\epsilon^*$
below which the speed increases {\em and}
the cost decreases as $\epsilon\downarrow 0$.
This suggests that, if such a protocol were
to be implemented in practice, it is the
relatively smaller values of $\epsilon$
that would be most effective in the long
run. 

\subsection{The continuous circle}
\label{s:cont_res}

Now we turn to the problem of $m=2$ walkers on the continuous
circle of circumference $N$, 
moving with constant speed $v$ and changing directions at rate $r$.
Theorem~\ref{thm:Cspeed} gives the natural continuous analog
of the discrete-time result in~Theorem~\ref{thm:speed2}.

\begin{theorem}[{\bf Message speed}]
\label{thm:Cspeed}
In the case of $m=2$ walkers, for any initial state,
the long-term average speed of the message is:
$$s:=v\pi(D_1(I_1))=\lim_{t\to\infty}\frac{1}{t}\int_{0}^{t}vD_s(I_s)ds
=\frac{v^2}{2v+rN},
\;\;\;\mbox{a.s.}$$
\end{theorem}

\noindent
Note that the speed $s=s(N,v,r)$ is always no greater than
$v/2$, as in the discrete case. Also observe that,
as would be expected, $s=s(N,v,r)$ is decreasing in the
circumference length~$N$ and increasing in 
the walker speed~$v$. Moreover, $s$ is also decreasing 
in the reversal rate $r$.

\medskip

Theorem~\ref{thm:Cspeed} answers
question~$(ii)$ of Section~\ref{s:problemC}.
The answer to question~$(i)$,
given below,
is an immediate consequence of Theorem~\ref{thm:Cspeed}.

\begin{corollary}[{\bf Message direction}]
\label{cor:Cdirection}
In the case of $m=2$ walkers, for any initial state $\Phi_0=\phi\in\Sigma$,
the steady state probability that the message moves
in the clockwise direction is:
$$P_\pi(D_1(I_1)=+1)=\lim_{t\to\infty}P_\phi(D_t(I_t)=+1)
=\frac{s+v}{2v}=\frac{3v+rN}{2(2v+rN)}.$$
\end{corollary}

\noindent
In our final result we determine
the long-term average number of jumps 
$c=c(N,v,r)$ per unit time. 

\begin{theorem}[{\bf Transmission cost}]
\label{thm:Ccost2}
In the case of $m=2$ walkers, for each time $t>0$ let $M_t$ denote
the (random) number of times the message jumps from one walker
to the other up to time $t$.
Then, for any initial state,
the long-term average cost of message transmissions is:
$$c:=\lim_{t\to\infty}\frac{M_t}{t}
=\frac{rv}{2v+rN},
\;\;\;\mbox{a.s.}$$
\end{theorem}

Observe that the cost $c=c(N,v,r)$ 
is naturally increasing in $v$ and 
decreasing in $N$.
But, unlike in the discrete case, 
$c=c(N,v,r)$ is monotonically increasing in $r$.
Again we also observe that there is 
a simple, scale-free relationship
between the asymptotic speed and 
cost, $c(N,v,r)=(r/v) \times s(N,v,r)$:
In the 
long-run, the message travels a 
(clockwise) distance of $v/r$ units
between successive jumps.

A comparison of the results of Theorems~\ref{thm:Cspeed}
and~\ref{thm:Ccost2} with their discrete-time analogs is
perhaps informative.
Consider a large discrete circle of size $N\gg 1$ 
and a small rate of direction updates $\epsilon\approx0$.
Then, noting that the circumference $N_c$ of the continuous 
circle corresponds to $2N$ in the discrete case (since,
because $N$ is odd, that
is the number of steps required for two walkers
starting in the same location and moving in opposite
directions to meet again),
we have the following scaling limit.
Taking the speed $v=1$ in the continuous case, 
and the rate $\epsilon$ in the discrete case to be
such that 
$2N\epsilon= N_c r$,
passing to the continuous limit 
$N\to\infty$
we obtain,
$$s(N,\epsilon)\to \frac{1}{2+rN_c}=s(N_c,v,r)
\;\;\;\mbox{and}\;\;\;
\frac{r}{\epsilon}c(N,\epsilon)\to \frac{r}{2+rN_c}=c(N_c,v,r).
$$

Finally we note that 
the parameters of the problem 
define the dimensionless quantity 
$\alpha =rN/v>0$, and dimensional analysis alone 
(meaning, speeding up time by a constant factor,
or dilating space by a constant factor) 
shows that $s=v f(\alpha)$ and $c=r g(\alpha)$, for 
suitable functions $f,g$. In this light, our
results can be interpreted as showing
that $f(\alpha)=g(\alpha)=1/(2+\alpha)$,
for all $\alpha>0$.


\section{Proofs}
\label{s:proofs}

\subsection{The discrete circle}

\noindent
{\sc Proof of Theorem~\ref{thm:speed2}. }
First,
consider the reduced
chain,
$$\bfPsi=\{\Psi_t=(Y_t=X_t(1)-X_t(2),D_t(1),D_t(2),I_t)\;;\;t\geq 0\},$$
where the differences
$Y_t=X_t(1)-X_t(2)$ are taken modulo $N$.
Clearly $\bfPsi$ is irreducible and aperiodic
on the corresponding reduced state space
$\Sigma_{\psi}$ consisting of 
all configurations,
of the form,
$$(y,d,d',i)\in
S\times\{+1,-1\}^2\times\{1,2\},$$
except $(0,+1,-1,2)$ and $(0,-1,+1,1)$.
Let $\pi_{\Psi}$ denote the unique invariant measure
of $\bfPsi$.
The limit in the theorem
exists a.s.\ by ergodicity; in order to 
compute its actual value,
we define the following regeneration time,
\be
T=\inf\{t\geq 1 \;;\;Y_t=0
\;\mbox{and}\;D_t(1)\neq D_t(2)\},
\label{eq:timeT}
\ee
and we consider
two special states of $\bfPsi$:
$\psi_1=(0,+1,-1,1)$ and 
$\psi_2=(0,-1,+1,2)$. Let $\nu$
denote the probability measure on
$\Sigma_{\psi}$ given by,
\be
\nu=\frac{1}{2}\delta_{\psi_1}+\frac{1}{2}\delta_{\psi_2}.
\label{eq:measnu}
\ee
Then $T$ is indeed a regeneration time for $\nu$ in the
sense that, with
$\Psi_0\sim\nu$, we also have
$\Psi_T\sim\nu$. We will use the following
general version of Kac's formula; cf.~\cite[Corollary~2.24]{aldous-fill:book}.

\begin{lemma}
\label{lem:kac}
For any function $f:\Sigma_{\psi}\to\RL$ and any regeneration time $T$ for $\nu$:
$$
E_\nu\left[\sum_{t=0}^{T-1}f(\Psi_t)\right]=
E_\nu(T) \pi_\Psi(f).$$
\end{lemma}

\noindent
To apply Lemma~\ref{lem:kac}, we first compute $E_\nu(T)$:

\begin{lemma}
\label{lem:ET2}
$E_\nu(T)=2N.$
\end{lemma}


\noindent
{\sc Proof. }
Consider the (further restricted) chain 
$\bfUpsilon=\{\Upsilon_t=(Y_t,D_t(1),D_t(2))\;;\;t\geq 0\}$
on the state space $S\times\{+1,-1\}^2$,
and note that its unique invariant measure
$\rho$ is uniform. Write, $D=\{(0,+1,-1),(0,-1,+1)\},$
let $\rho_D$ denote the measure $\rho$ conditioned
on $D$, and let,
$$T_D^+
=\inf\{t\geq 1\;;\;\Upsilon_t\in D\},$$
so that, in fact,
$T_D^+=T$. Then, by
Kac's formula \cite{aldous-fill:book},
we have,
$$E_\nu(T)=E_{\rho_D}(T_D^+)=\frac{1}{\rho(D)}=\frac{4N}{|D|}=2N,$$
as claimed.
\qed

The central step in the proof of the theorem
is an application of Lemma~\ref{lem:kac} with
$f(\Psi_t)=D_t(I_t)$, which, combined with
Lemma~\ref{lem:ET2} gives us
that
$s=\pi(D_1(I_1))=\pi_\Psi(D_1(I_1))$ can be
expressed as,
$$s
=
	\frac{1}{2N} E_\nu\left[\sum_{t=0}^{T-1}D_t(I_t)\right]
=
	\frac{1}{4N}
	E_{\psi_1}\left[\sum_{t=0}^{T-1}D_t(1)\right]
	+
	\frac{1}{4N}
	E_{\psi_2}\left[\sum_{t=0}^{T-1}D_t(2)\right]
=
	\frac{1}{2N}
	E_{\psi_1}\left[\sum_{t=0}^{T-1}D_t(1)\right],
$$
where the sums above (and in what follows) correspond
to addition over $\IN$ (as opposed to modulo $N$ addition
over $S$), and where the second equality follows from the fact that,
by the definition of $T$, the message is with walker~1 up to time
$T-1$.
Therefore, writing $X^*_{t+1}(j)=X^*_t(j)+D_t(j)$,
for $j=1,2$, $t\geq 1$, with $X^*_0(j)=X_0(j)$, for $j=1,2$, we have,
\ben
s
&=&
	\frac{1}{2N}E_{\psi_1}(X^*_T(1))\\
&=&
	\frac{1}{2N}E_{\psi_1}\left(\frac{X^*_T(1)-X^*_T(2)}{2}\right)
	+\frac{1}{2N}E_{\psi_1}\left(\frac{X^*_T(1)+X^*_T(2)}{2}\right)\\
&=&
	\frac{1}{2N}E_{\psi_1}\left(\frac{X^*_T(1)-X^*_T(2)}{2}\right),
\een
where we noted that 
$E_{\psi_1}(X^*_T(1)+X^*_T(2))$
is zero by symmetry,
since the two walkers start off in
opposite directions.

Now write, $A=\{(0,+1,+1),(0,+1,-1),(0,-1,+1),(0,-1,-1)\},$
and let $T_A^+$ denote the first time when the two walkers
meet,
$$T_A^+
=\inf\{t\geq 1\;;\;\Upsilon_t\in A\}
=\inf\{t\geq 1\;;\;Y_t=0\},$$
so that $T_A^+$ can be expressed in terms of 
either $\bfPsi$ or $\bfUpsilon$.
We observe that, at time $T_A^+$,
either the two walkers decide to go in 
opposite directions, in which case
$T_A^+=T$, or they continue moving together
until they choose opposite directions, in
which case the difference of their locations
$X^*_t(i)$ stays constant; therefore,
$$s=
	\frac{1}{2N}E_{\psi_1}
\left(\frac{X^*_{T_A^+}(1)-X^*_{T_A^+}(2)}{2}\right).$$

Since the last expectation above is conditioned
on the two walkers starting from the same
position, in opposite directions, and with
the first one moving in the positive (clockwise)
direction, there are exactly two possible scenarios
for their first meeting time $T_A^+$: In the 
first scenario,
at time $t=T_A^+-1$ walker~1 is two steps ``ahead''
in the clockwise direction of walker~2 (as they 
are, e.g., at time $t=1$). In this case, we will
necessarily have
$X^*_{T_A^+}(1)-X^*_{T_A^+}(2)=0$. We call this 
event $C$. In the second scenario, the relative
positions of the two walkers at time $t=T_A^+-1$
will be reversed,
which necessarily means that the first walker 
travelled a whole circle ``around''
the second one before they met, so that
(since $N$ is odd)
on $C^c$, we must have 
$X^*_{T_A^+}(1)-X^*_{T_A^+}(2)=2N$.
Therefore,
\be
s
=\frac{1}{2N}\left[\frac{0}{2}\cdot P_{\psi_1}(C)+
\frac{2N}{2}\cdot P_{\psi_1}(C^c)\right]
=\frac{1}{2}P_{\psi_1}(C^c).
\label{eq:almost2}
\ee
Finally we compute the probability of the event $C$:

\begin{lemma}
\label{lem:pround}
$P_{\psi_1}(C^c)=\frac{1-\epsilon}{1+\epsilon(N-2)}.$
\end{lemma}

\noindent
{\sc Proof. }
Here we consider the chain
$\bfUpsilon^*=\{\Upsilon^*_t=(Y^*_t,D_t(1),D_t(2))\;;\;t\geq 0\}$
on $\Sigma^*=\IN\times\{+1,-1\}^2$,
where 
$Y^*_t=X^*_{t}(1)-X^*_{t}(2)$. Note that, for the
state $u_1:=(0,+1,-1)$, the initial condition
$\Upsilon^*_0=u_1$ corresponds to $\Psi_0=\psi_1$. 

\begin{figure}[ht!]
\centerline{\includegraphics[width=6.0in]{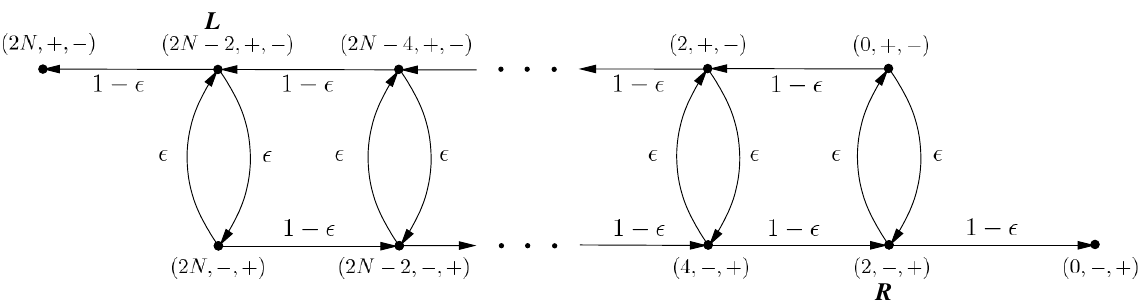}}
\caption{Evolution of the trace of the chain
$\bfUpsilon^*$ on $\{0,2,4,\ldots,2N\}\times\{(+1,-1),(-1,+1)\}$.}
\label{f:trace2}
\end{figure}

We will only need to examine the evolution of $\bfUpsilon^*$
until time $t=T_A^+$, which, since $N$ is odd,
can equivalently be expressed as,
$$T_A^+ =\min\{t\geq1\;;\;Y_t^*=0\pmod{2N}\},$$
and the same argument as in the last paragraph
before the statement of the lemma shows that,
given $\Upsilon^*_0=u_1$,
the only two possible values of $Y^*_{T^+_A}$ are~0
and~$2N$, on $C$ and on $C^c$, respectively.
Therefore, letting,
\ben
T_R&=&\min\{t\geq1\;;\;Y_t^*=0\},\\
\mbox{and}\;\;T_L&=&\min\{t\geq1\;;\;Y_t^*=2N\},
\een
we have that $T_A^+=\min\{T_L,T_R\}$ and
$P_{\psi_1}(C^c)=P_{\psi_1}(T_L<T_R)$;
cf.~Figure~\ref{f:trace2}.

In fact, for this computation it will suffice to consider
the {\em trace} of $\bfUpsilon^*$ on the set,
\ben
\Sigma^t:=\{0,2,4,\ldots,2N\}\times\{(+1,-1),(-1,+1)\}
\subset\Sigma^*;
\een
cf.~\cite[Example~1.4.4.]{norris:book}.
The evolution of this Markov chain is fairly simple
and its transition probabilities are easy to compute;
e.g., the probability of the transition from 
$(0,+1,-1)$ to $(2,+1,-1)$ is equal to,
$$(1-\epsilon)^2
+\epsilon(1-\epsilon)\frac{1}{2}
+\epsilon(1-\epsilon)\frac{1}{2}=
1-\epsilon.
$$
The first term above corresponds to the case when
the two walkers both maintain their original directions
after their first step;
the second term corresponds to the case when only the first 
walker changes direction, after which they keep moving at
a distance two apart, until one of them changes direction
again and they either reach the state $(2,+1,-1)$ or 
the state $(2,-1,+1)$, each having probability 1/2 by
symmetry; and the third term corresponds to the case 
when only the second walker changes direction after their
first step, and its value is the same as the second term
again by symmetry.
The remaining transition probabilities can be 
similarly computed; see~Figure~\ref{f:trace2}.

Finally, for every 
state $u\in\{0,2,4,\ldots,2N\}\times\{(+1,-1),(-1,+1)\}$
we define $h(u)=P_u(T_L<T_R)$, so that
$h(u_1)=P_{\psi_1}(C^c)$.
Writing $L$ and $R$ for the states
$(2N-2,+1,-1)$ and $(2,-1,+1)$, respectively,
we have $h(L)=1$, $h(R)=0$, and in
fact it is easy to see that the
one-step conditional expectation of $h$
given any state $u\neq (2N,+1,-1)$ or $(0,-1,+1)$, 
is equal to $h(u)$. This relationship can be expressed
as a simple recursion: Letting $f(k)=h(2k,+1,-1)$
and $g(k)=h(2k+2,-1,+1)$, 
we have,
$$\begin{cases}
f(k)=(1-\epsilon)f(k+1)+\epsilon g(k),&\mbox{for}\;0\leq k\leq N-1,\\
g(k+1)=(1-\epsilon)g(k)+\epsilon f(k+1),&\mbox{for}\;-1\leq k\leq N-2,\\
g(0)=0\;\mbox{and}\;f(N-1)=1.
\end{cases}
$$
Adding the first two equations above
shows that $f(k)-g(k)$ is a constant,
say $A$, independent of $k$,
and substituting this in the recursion
for $g$ gives $g(k)=A\epsilon k/(1-\epsilon)$.
Similarly solving for $f$ we obtain,
$f(k)=A+A\epsilon k/(1-\epsilon)$,
and from the boundary values we can
solve for $A$ to get,
$A
=(1-\epsilon)/(1+\epsilon(N-2))$.
Therefore,
$$P_{\psi_1}(C^c)=h(u_1)=f(0)=A
=(1-\epsilon)/(1+\epsilon(N-2)),$$
as claimed.
\qed

\noindent
Combining~(\ref{eq:almost2}) with
the result of Lemma~\ref{lem:pround}
completes the proof of the theorem.
\qed



\medskip

\noindent
{\sc Proof of Theorem~\ref{thm:cost2}. }
Recall the ergodic chain $\bfPsi$ defined in the 
beginning of the proof of Theorem~\ref{thm:speed2}.
Write $\Sigma_\psi$ for its state space,
$\pi_\Psi$ for its unique invariant measure,
and let $\PB$ denote its transition kernel,
$\PB(\psi,\psi')=\Pr(\Psi_{t+1}=\psi'|\Psi_t=\psi)$,
$\psi,\psi'\in\Sigma_\psi$.
Now consider the bivariate chain
$\bftPsi=\{\tilde{\Psi}_t=(\Psi_t,\Psi_{t+1})\;;\;t\geq 0\}.$
Then $\bftPsi$ is also ergodic, with unique invariant 
measure,
$$\tilde{\pi}(\psi,\psi')=\pi_\Psi(\psi)\PB(\psi,\psi'),$$
for every state $(\psi,\psi')$ of $\bftPsi$.
Therefore, the limit in the statement indeed exists a.s.,
and it equals,
$$
c:=P_{\pi}(I_{2}\neq I_1)
=
\tilde{\pi}\Big(\IND_{\{I_{2}\neq I_1\}}\Big)
=
\tilde{\pi}(B),
$$
where $B$ consists of the following 8 states,
\ben
B=\Big\{\hspace{-0.1in}
&\hspace{-0.2in}\Big( (0,+1,+1,1),(0,-1,+1,2)\Big),&\Big( (0,-1,-1,1),(0,-1,+1,2)\Big),\\
&\hspace{-0.2in}\Big( (0,+1,+1,2),(0,+1,-1,1)\Big),&\Big( (0,-1,-1,2),(0,+1,-1,1)\Big),\\
&\hspace{-0.2in}\Big( (2,-1,+1,1),(0,-1,+1,2)\Big),&\Big( (2,-1,+1,2),(0,+1,-1,1)\Big),\\
&\Big( (-2,+1,-1,1),(0,-1,+1,2)\Big),&\Big( (-2,+1,-1,2),(0,+1,-1,1)\Big)
\Big\},
\een
and where, with a slight abuse of notation,
the negative values of the $Y_t$ variables above are
again interpreted modulo $N$.

In order to compute the actual value
of $c=\tilde{\pi}(B)$, we first observe 
that,
\ben
\tilde{\pi}(B)
\!&=&\!\!
	\Big(\pi_\Psi(0,+1,+1,1)+\pi_\Psi(0,-1,-1,1)
	+\pi_\Psi(0,+1,+1,2)+\pi_\Psi(0,-1,-1,2)\Big)\epsilon
	(1-\epsilon)\\
\!&&\!
	+\Big(\pi_\Psi(2,-1,+1,2)+\pi_\Psi(-2,+1,-1,1)\Big)
	\epsilon^2\\
\!&&\!
	+\Big(\pi_\Psi(2,-1,+1,1)+\pi_\Psi(-2,+1,-1,2)\Big)
	(1-\epsilon)^2\\
\!&=&\!\!
	\frac{1}{2N}\epsilon
	(1-\epsilon)
	+\Big(\pi_\Psi(-2,+1,-1,1)-\pi_\Psi(2,-1,+1,1)+\frac{1}{4N}\Big)
	\epsilon^2\\
\!&&\!
	+\Big(
	\pi_\Psi(2,-1,+1,1)
	-\pi_\Psi(-2,+1,-1,1)+\frac{1}{4N}
	\Big)
	(1-\epsilon)^2,
\een
where we used the fact that the invariant
distribution of $(X_t(1),X_t(2),D_t(1),D_t(2))$ is uniform,
which implies that
$\pi_\Psi(y,d,d',1)+\pi_\Psi(y,d,d',2)=1/(4N)$,
for any $y\in S$ and $d,d'\in\{+1,-1\}$.
Simplifying,
\be
\tilde{\pi}(B)
=\frac{1}{4N}+(1-2\epsilon)
	[\pi_\Psi(2,-1,+1,1)
	-\pi_\Psi(-2,+1,-1,1)].
\label{eq:pitB}
\ee

To compute the difference of the two probabilities in~(\ref{eq:pitB}),
recall the definition of the regeneration time $T$
and the measure $\nu$ in~(\ref{eq:timeT}) and~(\ref{eq:measnu}),
respectively. By Lemma~\ref{lem:kac}, for any
state $\psi\in\Sigma_\psi$ of the form $\psi=(y,d,d',1)$, we have,
$$E_\nu(T)\pi_\Psi(\psi)=
E_\nu\left[\sum_{t=0}^{T-1}\IND_{\{\Psi_t=\psi\}}\right]
=
\frac{1}{2}
E_{\psi_1}\left[\sum_{t=0}^{T-1}\IND_{\{\Psi_t=\psi\}}\right]
=
\frac{1}{2}
E_{u_1}\left[\sum_{t=0}^{T-1}\IND_{\{\Upsilon^*_t=(y,d,d')\}}\right],$$
where the chain $\bfUpsilon^*$ was defined in the proof
of Lemma~\ref{lem:pround} and $u_1=(0,+1,-1)$ as before.
Therefore, substituting this twice in~(\ref{eq:pitB})
and recalling the discussion of the evolution 
of $\bfUpsilon^*$ until time $T$ from the proof 
of Lemma~\ref{lem:pround}, we have,
\be
c=\tilde{\pi}(B)=\frac{1}{4N}+\frac{(1-2\epsilon)}{4N}
E_{u_1}\left[\sum_{t=0}^{T-1}
\Big(
\IND_{\{\Upsilon^*_t=(2,-1,+1)\}}
-\IND_{\{\Upsilon^*_t=(2N-2,+1,-1)\}}
\Big)\right],
\label{eq:easysum}
\ee
where we also 
used the result of Lemma~\ref{lem:ET2}.
By the definition of $T$, the value of the
sum inside the last expectation above
is either $0-1$ or $1-0$, and the corresponding
probabilities can be found by looking at the trace
of the chain $\bfUpsilon^*$ 
on the set $\Sigma^t$,
as defined 
in the proof of Lemma~\ref{lem:pround}.
Indeed, referring to Figure~\ref{f:trace2},
and in the notation of the proof of Lemma~\ref{lem:pround},
the case $1-0$ has probability
$1-P_{u_1}(T_L<T_R)$,
whereas the case $0-1$ has probability
$P_{u_1}(T_L<T_R)$. So~(\ref{eq:easysum})
becomes,
$$c=\tilde{\pi}(B)=\frac{1}{4N}+\frac{(1-2\epsilon)}{4N}
[1-2P_{u_1}(T_L<T_R)],
$$
and now substituting the result of Lemma~\ref{lem:pround},
$P_{u_1}(T_L<T_R)=\frac{1-\epsilon}{1+\epsilon(N-2)}$,
and simplifying, yields precisely the claimed result.
\qed

\subsection{The continuous circle}

\noindent
{\sc Proof of Theorem~\ref{thm:Cspeed}. }
We begin with some simple notation.  
Let,
\begin{eqnarray*}
D
&=&
	\{(z,z): z\in S\}\subset S^2\\
F
&=&
	\{(z_1,z_2,d_1,d_2,i)\in\Sigma : z_1=z_2,\;d_1d_2=-1\}
	\subset\Sigma,
\end{eqnarray*}
and note that, by the definition of $\Sigma$,
for any $(z_1,z_2,d_1,d_2,i)\in F$
we can always take, without loss of generality, $d_i=+1$.

The limit in the statement of the theorem exists a.s.\ by 
Theorem~\ref{thm:ergodic}; 
in order to compute its value we first
define the stopping time,
\begin{equation}\label{regent}
T=\inf\big\{t>0:\ \Phi_t\in F\big\}.
\end{equation}
Let $\nu$ denote 
the uniform probability measure on $F$.
Then, $T$ is a regeneration time for $\nu$, in the sense that,
if $\Phi_0\sim \nu$, then $\Phi_T\sim\nu$, as well.
We go on to compute $\pi(D_1(I_1))$ using the following
natural continuous-time generalization of Kac's formula,
proved in the Appendix.

\begin{lemma}
\label{renewal} 
If $T$ is a regeneration stopping 
time for $\nu$ with $E_\nu(T)<\infty$, 
then for any bounded measurable function $f:\Sigma\to\RL$ we have,
\[
E_\nu\left(\int_0^T f(\Phi_s) ds
\right)=E_\nu(T) \pi(f).
\]
\end{lemma}

\noindent
First, we compute the expectation 
of the regeneration time~$T$,
conditional on the initial state $\Phi_0=\phi$
being in $F$.

\begin{lemma}\label{exptime}
For any initial state $\phi\in F$, we have:
$E_\phi(T)=\frac{N}{v}.$
\end{lemma}

\noindent
{\sc Proof. } For any state
$\phi=(x,d,i)=(x_1,x_2,d_1,d_2,i)\in \Sigma$,
write,
$$\delta(x):=x_1-x_2\;(\text{mod}\, N)\in [0,N).$$
We can compute $E_\phi(T)$ using simple tools from the
potential theory of Markov processes. To that end, we will
construct a function $H:\Sigma\to \RL$ that formally 
satisfies $LH(\phi)=-1$, for all $\phi\in \Sigma\setminus F$. 
Note that such a function would not be in the domain of the generator,
as can be easily seen by integrating both sides with respect 
to the invariant measure. Nevertheless, if we start the process 
from $\Phi_0=\phi\notin F$,
then it can be checked that $\{H(\Phi_t)+t\}$ is a martingale up 
to time $T$. 

It is not hard to find an explicit solution 
to $LH(\phi)=-1$ for all $\phi\notin F$. Indeed, 
let, for $\phi\notin F$,
$$H(\phi):=\left(\frac{N-2\delta(x)}{4v}\right)\big(d_1-d_2\big)
+\frac{1+d_1d_2}{4r}+\frac{r\,\delta(x)\big(N-\delta(x)\big)}{2v^2}, $$
and for $\phi\in F$ as,
\[
H(\phi):=-\frac{N}{2v}.
\]
Recalling the form of the generator $L$ from~(\ref{eq:generator}),
it is straightforward to verify that $H$ satisfies,
$
LH(\phi)=-1,
$
for all $\phi\notin F$,
and that $H$ is discontinuous across $F$, in that,
\[
\lim_{\delta(x)\downarrow 0} H(x,d,i)-\lim_{\delta(x)\uparrow N}H(x,d,i)
=\frac{N}{2v}\big(d_1-d_2\big).
\]
For $\phi\not\in F$, an 
application of the optional stopping theorem 
for the martingale $\{H(\Phi_t)+t\}$ gives,
\begin{equation}\label{outside}
E_{\phi}(T)=H(\phi)+\frac{N}{2v}.
\end{equation}
On the other hand, for $\phi\in F$, the Markov property gives,
\[
E_{\phi}(T)=E_{\phi}\big[T;\ T\leq t\big]+E_{\phi}\big[E_{\Phi_t}(T);\ T>t\big],
\]
for all $t>0$.
Since, as $t\to 0$, we have
$P_{\phi}(T\leq t)\to 0$ and $H(\Phi_t)
\to
N/(2v)$,
$P_{\phi}$-a.s., 
for all $\phi\in F$, 
letting $t\to 0$ in the preceding equation
and recalling \eqref{outside} 
completes the proof.
\qed

Now let 
$d^*:=(+1,-1)$, 
$\phi^*:=(0,0,d^*,1)\in F$,
and note that,
for all $x\in D$,
$$E_{(x,d^*,1)}\left[\int_0^T D_t(I_t)\, dt\right]
=E_{(x,-d^*,2)}\left[\int_0^T D_t(I_t)\, dt\right]
=E_{\phi^*}\left[\int_0^T D_t(1)\, dt\right].
$$
Combining this, with an application 
of Lemma~\ref{renewal} with $f(\Phi_t)=D_t(I_t)$, 
and with Lemma~\ref{exptime}, gives,
\ben
s=\frac{v^2}{N}E_\nu\left[\int_0^TD_t(I_t)\, dt\right]
=\frac{v^2}{N}E_{\phi^*}\left[\int_0^T D_t(1)\, dt\right].
\een
Therefore, writing $X_t^*(i)$
for the total clockwise displacement distance travelled
by walker $i=1,2,$ up to time $t\ge 0$, 
so that $X_t^*(i)=v\int_0^t D_t(i)\, dt$, we have,
\ben
s&=&\frac{v}{N} E_{\phi^*}\big[X_T^*(1)\big]\\
&=&\frac{v}{2N}
E_{\phi^*}\big[X_T^*(1)+X_T^*(2)\big]
+ \frac{v}{2N}E_{\phi^*}\big[X_T^*(1)-X_T^*(2)\big]\\
&=&\frac{v}{2N}E_{\phi^*}\big[X_T^*(1)-X_T^*(2)\big],
\een
where we used the fact that, since the two walkers start in 
opposite directions, we have
$E_{\phi^*}[X_T^*(1)+X_T^*(2)]=0$
by symmetry.

There are exactly two scenarios for the first meeting of the two 
walkers starting from $F$ with $D_0(I_0)=1$: 
In the first one, they meet with directions 
that are opposite to the ones they started with. In 
this case, we necessarily have $X_T^*(I_0)-X_T^*(I'_0)=0$, where 
$I'_0$ denotes the complementary index to $I_0$,
i.e., $I_0'=3-I_0$.
We call this event $C$.
In the second scenario, corresponding to event $C^c$, the directions of the 
walkers are the same as the ones they started with, which necessarily 
means that $X_T^*(I_0)-X_T^*(I'_0)=N$. Therefore,
the speed $s$ is:
\[
s=\frac{v}{2}P_{\phi^*}(C^c).
\]
The proof of the theorem
is completed by an application
of Lemma~\ref{eventC} below.
\qed

\begin{lemma}\label{eventC}
For any initial state $\phi\in F$ of the form
$\phi=(x,1,-1,1)$, we have:
$$P_{\phi}(C^c)=\frac{2v}{2v+rN}.$$
\end{lemma}

\noindent
{\sc Proof. }
Consider the function $V:\Sigma\setminus F\to \RL$,
defined, for any state $\phi=(x,d_1,d_2,i)\notin F$ as,
\[
V(\phi):=\frac{r\delta(x)+v[1+\big(d_1-d_2\big)/2]}{rN+2v},
\]
where $\delta(x):=x_1-x_2\;(\text{mod}\, N)$, as before.
It is straightforward to verify that $LV(\phi)=0$, 
for every $\phi\notin F$. An application of the optional stopping 
theorem for
the martingale $\{V(\Phi_t)\}$ gives,
\[
P_{\phi}(C^c)=V(\phi),
\]
for all $\phi=(x,1,-1,1)$ with $x\notin D$.
The proof is concluded by repeating the same argument
as in the end of the proof of Lemma~\ref{exptime},
with the random variable $\IND_{C^c}$ in place of $T$,
and using the Markov property at time $t>0$ 
and letting $t\to 0$.
\qed

\noindent
{\sc Proof of Theorem~\ref{thm:Ccost2}. }
Using 
the same notation as in the proof of Theorem~\ref{thm:Cspeed},
we define, for each $n\geq 0$, the time
$T_n$ as the time of the $n$th return of $\mathbf{\Phi}$ to $F$, 
i.e., $T_0:=0$ and inductively, for all $n\geq 1$,
\[
T_{n+1}:=\inf\{t>T_n: \Phi_t\in F\}.
\]
The number of excursions around $F$ up to time $t\ge 0$ is then,
\[
N_t=\max\{n\geq 0:\ T_n\leq t\}.
\]
We also define the independent Bernoulli random variables 
$\{J_n;n\geq 1\},$ that take the value 1 exactly when 
the event $C$ occurs during the $n$th excursion around $F$.
Lemma~\ref{eventC} then implies that, for all $n\geq 2$,
\[
P_\phi(J_n=1)=\frac{rN}{2v+rN}.\]
In this notation, the total transmission cost $M_t$ up to time $t\ge 0$ 
is given by,
\[
M_t=\sum_{n=1}^{N_t}J_n.
\]
Since, by Lemma~\ref{exptime}, $N_t/t\to v/N$, a.s., 
as $t\to\infty$, we have,
\[
c:=\lim_{t\to\infty} \frac{M_t}{t}= \frac{v}{N}\times \frac{rN}{2v+rN}
=\frac{rv}{2v+rN},\;\;\;\mbox{a.s.},
\]
as claimed.
\qed


\appendix

\section{Appendix}

\subsection{Proof of Proposition~\ref{prop:ergodic0}}

Parts~$(i)$ and~$(ii)$ are immediate 
consequences of the more general result 
in Theorem~\ref{thm:ergodic}, established
below.
For~$(iii)$ it is easiest to work with the
infinitesimal generator $L_0$ of $\bfW$. 
Let $A_{L_0}$ denote the collection 
of all continuous functions 
$f:S^m\times\{-1,+1\}^m\to\RL$, such that
$f$ is continuously differentiable
in each $x(j)$, $1\leq j\leq m$.
Also, for any $m$-tuple of directions $d\in\{-1,+1\}^m$,
let $\sigma^jd$ be the same as $d$ but with
its $j$th coordinate having the opposite sign
from that of $d$,
$1\leq j\leq m$.
Then the action of $L_0$ on any $f\in A_{L_0}$ is,
$$L_0f(x,d)=\sum_{j=1}^m 
\left\{v d(j)\frac{\partial f}{\partial x(j)}(x,d)+r\left[
f(x,\sigma^jd)-f(x,d)
\right]
\right\}.
$$
Arguing as for the operator $L$ at the end of Section~\ref{s:problemC}
we get that the domain of $L_0$ is $A_{L_0}$.
It is now a simple computation to show that,
if $\pi_0$ denotes the uniform distribution
on 
$S^m\times\{-1,+1\}^m$,
then $\int L_0fd\pi_0=0$, for any $f\in A_{L_0}$,
so that $\pi_0$ is indeed invariant \cite[Theorem~3.37]{liggett:10}.
The uniqueness of $\pi_0$ follows from
Harris recurrence \cite{getoor:80}.
\qed

\subsection{Proof of Theorem~\ref{thm:ergodic}}

For the sake of simplicity, we assume that
all three parameters, $N$, $v$ and $r$, are equal to~1. 
A cursory examination of the proof below should immediately 
reveal that the general case only involves notational
modifications.

Given $\epsilon\in(0, 1/(3m))$ arbitrary, 
let $S_\epsilon\subset S^m$ denote the 
set,
$$S_\epsilon :=S^m\setminus 
\{x: d(x(j),x(k))<3\epsilon,\;\mbox{for some}\;j\neq k \},$$
where $d(\cdot,\cdot)$ denotes the usual distance on $S$, 
so that $d(x(j),x(k))$
is equal to the length of the shortest arc connecting points
$x(j)$ and $x(k)$.
Note that, since $\epsilon<1/(3m)$, $S_\epsilon$ has nonempty interior.
We also define the set
$\Sigma_\epsilon:=S_\epsilon \times\{-1, 1\}^m
\times\{1, 2,\ldots, m\}$,
and the measures
$\clL_\epsilon$ such that
$d \clL_\epsilon/d\clL^m=$
$\IND_{S_\epsilon}$ on $S^m$,
and 
$\mu:=\clL_\epsilon\times \kappa^m 
\times \kappa_m$ on $\Sigma$.
Note that $\clL_\epsilon$ and $\mu$
are supported on $S_\epsilon$
and $\Sigma_\epsilon$, respectively.

The main step in the proof is the following Doeblin-like
domination condition:

\begin{proposition}
\label{prop:doeblin}
Let $\epsilon\in(0, 1/(3m))$ arbitrary.
Then, for every $t\ge t_0:=2+6\epsilon$, every measurable $A\subset \Sigma$, 
and every initial state $\phi\in \Sigma$, we have,
\be
P_\phi(\Phi_t\in A)  \ge c \mu(A),
\label{eq:original}
\ee
with $c=(\epsilon^2 e^{-1-(3/2)t_0}/4)^m$.
\end{proposition}

Before giving the proof we make two simple observations. 
First, we will actually prove that, under the assumptions 
of the proposition,
\be
P_\phi(\Phi_t\in A)  \ge c'e^{-(3/2)tm} \mu(A),
\label{eq:modified}
\ee
with $c'=(\epsilon^2 e^{-1}/4)^m$.
Then~(\ref{eq:original}) follows for each $t\geq t_0$
by the Markov property:
$$
P_\phi(\Phi_t\in A) 
=E_\phi\big[P_{\Phi_{t-t_0}}\Phi_{t_0}\in A
\big]
\ge c\mu(A).$$
Second, it suffices to establish~(\ref{eq:modified})
for events $A$ of the form,
\be
A= \prod_{j=1}^m B_j
\times \prod_{j=1}^m \{d(j)\} \times\{1\},
\label{eq:eventA}
\ee
with each $d(j)\in\{-1, 1\}$, and 
each $B_j\subset S$.
The proof is based on the following construction.

For any two points $x,x'\in S$, we say that a walker 
{\em travels `clockwise' from $x$ 
to $x'$ in time $t$}, when the walker takes the following
steps. Let $x''=\min\{y\in(x,\infty):y=x'\;(\mbox{mod}\;1)\}$.
Then the walker, starting at $x$, travels counter-clockwise 
for time $z=(t+x-x'')/2$,
then switches directions and travels 
clockwise for time $t-z$, ending up at point $x''$.
Similarly, we say that a walker
{\em travels `counter-clockwise' from $x$ 
to $x'$ in time $t$}, when the reverse of
the above process occurs, so that, now,
$x''=\max\{y\in(-\infty,x):y=x'\;(\mbox{mod}\;1)\}$.

Let $A$ be as in~(\ref{eq:eventA}).
A simple strategy for the 
motion of the $m$ walkers that guarantees $\Phi_t\in A$ is 
the following:  Walker~1 travels clockwise from point $x(1)$ 
to some point in $B_1$ in time $t$, 
while all other walkers travel counter-clockwise from 
their initial positions $x(k)$ to some point in the
corresponding $B_k$ in time $t$. During the final segment 
of her travel, walker~1 will encounter each of the 
other walkers at least once, and she will be moving
clockwise while all of the other walkers will be moving
counter-clockwise. Therefore, 
the message will 
certainly be with walker~1 at time $t$.

\medskip

\noindent
{\sc Proof of Proposition~\ref{prop:doeblin}. }
Let $\epsilon\in(0, 1/(3m))$ 
and $\phi=(x_0,d_0,i_0)\in\Sigma$ arbitrary,
and without loss of generality take $A\subset\Sigma$ 
to be of the form~(\ref{eq:eventA}).

For any position $x\in S$, any 
pair of directions $d,d'\in\{-1,+1\}$, any time $t>0$,
and any measurable $C\subset S$, we will define
the event $M^+(x,d,C,d',t,\epsilon)$ that,
roughly speaking, will describe a scenario
in which 
a walker starts at $x$ with direction $d$,
and ends up in $C$ with direction $d'$ after time $t$.
To make this precise, let $e_0,e_1,e_2$ and $e_3$ be 
independent exponential random variables with mean 
equal to~1, and imagine a walker starting at time zero 
in position~$x$ with direction~$d$, taking
the following steps:
\begin{itemize}
\item $+e_0,-e_1,+e_2$, if $d=d'=1$ (i.e.,
the walker first moves clockwise for time
$e_0$, then counter-clockwise for time $e_1$,
and then clockwise again for time $e_2$);
\item $+e_0,-e_1,+e_2,-e_3$, if $d=1$, $d'=-1$;
\item $-e_1,+e_2$, if $d=-1$, $d'=1$;
\item $-e_1,+e_2,-e_3$, if $d=d'=-1$.
\end{itemize}
Now we can compute the location of
the walker at time $t$.
Write $E_0=e_0\IND_{\{d=1\}}$
and consider two cases:

If $d'=1$, and assuming that,
\be
\frac{t-1}{2}< e_1<\frac{t}{2},
\;\;\;e_2>t,
\;\;\;\mbox{and}
\;\;\;E_0\leq \epsilon,
\label{eq:cond1}
\ee
then at time $2e_1$ the walker 
has covered the distances $E_0$, $-e_1$,
and $e_1-E_0$, and therefore is at point $x$
and is moving clockwise. Moreover,
since $e_2>t>2e_1$, after the remaining $t-2e_1$
time units, the walker ends up in
position $x+t-2e_1$.

If $d'=-1$, and assuming that,
\be
E_0+e_1+e_2<t<E_0+e_1+e_2+e_3,
\label{eq:cond2}
\ee
then at time
$E_0+e_1+e_2$ 
the walker is located at $x+E_0-e_1+e_2$
and moving counter-clockwise. 
Therefore, at time $t$, i.e., 
after travelling counter-clockwise for an additional 
$t-(E_0+e_1+e_2)<e_3$ time units, the walker's final 
position is 
$x+E_0-e_1+e_2-(t-E_0-e_1-e_2)=x-t+2(E_0+e_2)$.

To summarize, under assumptions~(\ref{eq:cond1}) 
and~(\ref{eq:cond2}), the position (in $\RL$, not necessarily
in $S$), of the walker at time $t$ is,
\be
U:=
\begin{cases} x+t-2e_1, & \text{ if } d'=1,\\
x-t+2(E_0+e_2), & \text{ if } d'=-1.
\end{cases}
\label{eq:U}
\ee
The last ingredient we need for the formal definition
of $M^+$ is the following. Given $x$ and $C$, we write,
$$C_x:= (C\cap(x, 1])\cup (C\cap[0, x]+1) \subset (x, x+1],$$
so that $C_x$ is the same as $C$, but the points before $x$ 
have been pushed one unit later. Now we can formally
define:
$$M^+(x, d, C, d',t,\epsilon):=
\begin{cases} 
\{E_0\le \epsilon, e_2>t, U\in C_x\}, &\text{ if } d'=1,\\
\{E_0\le \epsilon, e_3>\epsilon, E_0+e_1+e_2\in (t-\epsilon, t),  U\in C_x\},
	 &\text{ if } d'=-1.
\end{cases}
$$
Note that both conditions~(\ref{eq:cond1}) and~(\ref{eq:cond2}) above
are satisfied on $M^+(x, d, C, d',t,\epsilon)$. We also define
the ``symmetric'' event
$M^-(x,d,C,d',t,\epsilon)$ as the event in which the 
reflection of the walker follows the path described 
by $M^+(-x,-d,-C,-d',t,\epsilon)$. Finally, let,
\ben
G:=\left\{
\begin{array}{l}
\text{The path of walker~$1$ up to time $t$ is in } 
	M^+(x_0(1),d(1),B_1,d'(1),t,\epsilon), \\
\text{and the path of each walker $j\ne 1$ up to $t$ is in } 
M^-(x_0(j),d(j),B_k,d'(j),t,\epsilon)
\end{array} 
\right\}.
\een

Now we claim that, for all $t>t_0:=2+6\epsilon$:
\be
G\subset \{\Phi_t\in A\}.
\label{eq:claim}
\ee
To see that~(\ref{eq:claim}) holds, again we consider two cases.
If $d'=1$, then time $t$ is reached during step $e_2$, because 
$E_0<\epsilon, e_1=(t+x-U)/2<t/2$ and $e_2>t$, so that $E_0+e_1<t<E_0+e_1+e_2$.
And if $d'=-1$, then time $t$ is reached during step $e_3$.
Therefore, a walker following $M^+(x, d, C, d',t, \epsilon)$ 
travels clockwise at least in the time interval 
$[t-(1+2\epsilon), t-\epsilon]$, which has length $1+\epsilon$. 
To see this, note that, if $d'=1$, then we 
have $e_1=(t-(U-x))/2>\epsilon\ge E_0$,
and the last part of the trajectory (i.e., the part of step $e_2$ 
until time $t$) 
takes time,
$$e_1-E_0+U-x=(t+U-x)/2-E_0\ge t/2-\epsilon\ge 1+2\epsilon,$$
while if
$d'=-1,$ then,
$$e_2=(U-x+t)/2-E_0\ge t/2-\epsilon\ge 1+2\epsilon,$$ 
and also $E_0+e_1+e_2\in(t-\epsilon, t)$.  Consequently, 
in the time interval 
$[t-(1+2\epsilon), t-\epsilon]$, walker~1 meets every other walker 
and thus gets 
the message. In the remaining time interval, $[t-\epsilon, t]$, the walkers do 
not meet again because each is at distance at most $\epsilon$ from their
final positions, and the distance between any two of these
final positions is at least $3\epsilon$.

Now,~(\ref{eq:claim}) implies that,
\ben
P_\phi(\Phi_t\in A)
&\geq&
	P_\phi(G)\\
&=&
	P_\phi\big( M^+(x_0(1),d(1),B_1,d'(1),t,\epsilon)\big)
	\prod_{j=2}^m
	P_\phi
	\big(M^-(x_0(j),d(j),B_k,d'(j),t,\epsilon)\big),
\een
and the bound in~(\ref{eq:modified}) and
hence the result of the proposition follow from the bound
in the lemma immediately below.
\qed

\begin{lemma}
\label{lem:doeblin}
In the notation and under the assumptions
of the above proof, for any 
position $x\in S$, and pair of directions
$d,d'\in\{+1,-1\}$, any measurable $C\subset S$,
and any initial state $\phi\in\Sigma$, we have,
$$
	P_\phi\big( M^+(x,d,C,d',t,\epsilon)\big)
	\geq c_\epsilon\clL_\epsilon(C),
$$
with $c_\epsilon=\frac{1}{4}e^{-1}\epsilon^2e^{-3t/2}$.
By symmetry, the same bound holds for $M^-$ in place
of $M^+$.
\end{lemma}

\noindent
{\sc Proof of Lemma~\ref{lem:doeblin}. }
When $d'=1$, the random variable $U$ defined in~(\ref{eq:U}) has density 
$f_U(s)=(1/2) e^{-(x+t)/2} e^{s/2}\IND_{\{s<x+t\}}$, so that,
$$P_\phi(M^+(x, d, C, d',t, \epsilon))=(1-e^{-\epsilon}\IND_{\{d=1\}})
e^{-t}\frac{1}{2} e^{-(x+t)/2} \int_{C_x} e^{s/2}\, ds\ge 
\frac{1}{4} \epsilon e^{-1/2} e^{-3t/2}\clL_\epsilon(C),$$
where we used the elementary inequality,
$1-e^{-\epsilon}\ge\epsilon/(1+\epsilon)\ge \epsilon/2$,
$\epsilon\in[0, 1]$.

When $d'=-1$, we consider two cases.
If $d=-1$, then the joint density of $(U, V):=(x-t+2e_2, e_1+e_2)$ 
is,
$$f_{U, V}(u, v)=\frac{1}{2}e^{-v} \IND_{\{u>x-t\}} \IND_{\{2v-u>t-x\}},$$
and thus,
\ben 
P_\phi(M^+(x, d, C, d',t, \epsilon))
&=&
	\Pr(U\in C_x, V\in (t-\epsilon, t))\\
&=&
	\frac{1}{2}\int_{C_x} \int_{t-\epsilon}^{t}e^{-v}\, dv\, du\\
&\ge&
	\frac{1}{2}\epsilon e^{-1} e^{-t}\clL_\epsilon(C).
\een
Finally, when $d=1$, the joint density of
$(U, V, W):=(x-t+2(e_0+e_2), e_0+e_1+e_2, e_0)$ is,
$$f_{U, V, W}(u, v, w)=\frac{1}{2}e^{-v}\IND_{\{2v-u>t-x\}}
\IND_{\{0<w<(u+t-x)/2\}},$$
and thus,
$$
P_\phi(M^+(x, d, C, d',t, \epsilon))=\frac{1}{2}\int_0^{\epsilon}\int_{C_x} \int_{t-\epsilon}^{t}e^{-v}\, dv\, du\, dw\ge \frac{1}{2}\epsilon^2 e^{-t}\clL_\epsilon(C).
$$
Combining the three bounds derived gives the required result.
\qed

We are now in a position to establish the four claims 
of Theorem~\ref{thm:ergodic}. If $\psi(A)>0$ 
for some $A\subset \Sigma$, then 
$\mu(A)=\psi(A\cap \Sigma_\epsilon)>0$
for all $\epsilon>0$ small enough, so
Proposition~\ref{prop:doeblin} implies that
$\bfPhi$ is $\psi$-irreducible \cite{meytwe93a}. Moreover,
it implies that the state space $\Sigma$ itself
is small \cite{meytwe93a}, hence it is petite,
so that $\bfPhi$ is aperiodic \cite[Eq.~(10)]{dowmeytwe}. This 
establishes~$(i)$.

Since $\Sigma$ is small, the drift condition $(\clD_T)$
of \cite{dowmeytwe} holds with Lyapunov function $V\equiv 1$,
and \cite[Theorem~5.2]{dowmeytwe} implies that
$\bfPhi$ is uniformly (exponentially) ergodic.
This implies~$(iii)$ by the definition of $V$-uniform
ergodicity \cite[Eq.~(11)]{dowmeytwe}. In particular, $\bfPhi$
is positive Harris recurrent, hence it has a unique invariant
probability measure \cite{getoor:80}, giving~$(ii)$.
Finally, the ergodic theorem of part~$(iv)$ is 
a standard consequence of positive recurrence
of strong Markov processes, see, e.g.,~\cite[Theorem~5.1]{maruyama:59}
or \cite[Proposition~3.7]{asmussen:book}.
\qed

\subsection{Proof of Lemma~\ref{renewal}}
If  $P_\nu(T=0)=1$, the claim is trivial. 
Suppose now $0<E_\nu(T)<\infty$,
and consider the following occupation 
measure $\zeta$ on $\Sigma$: For any measurable $A$,
\[
\zeta(A)=E_\nu\left(\int_0^T {\mathbb I}_A(\Phi_s) ds\right).
\]
Observe that $\zeta$ integrates measurable functions on $\Sigma$ as,
\[
\int f\, d\zeta=E_\zeta\left(\int_0^T f(\Phi_s) ds\right).
\]
Let now $g\in {\cal D}(L)$. By an application of the optional stopping theorem for the martingale  $\{g(\Phi_t)-\int_0^t L g(\Phi_s) ds\}$
at the stopping time $T$, we get,
\[
E_\nu\big[g(\Phi_T)\big]
-E_\nu\left(\int_0^T Lg(\Phi_s) ds\right)
=E_\nu\big[g(\Phi_0)\big].
\]
As $T$ is a regeneration time for $\nu$, 
we have that,
$E_\nu\big[g(\Phi_T)\big]
=E_\nu\big[g(\Phi_0)\big]$,
and hence,
\[
\int Lg\, d\zeta=E_\nu\left(\int_0^T Lg(\Phi_s) ds\right)=0.
\]
Since this holds for all $g\in{\cal D}(L)$, 
the normalized occupation 
measure $\bar{\zeta}:=\zeta/\zeta(\Sigma)=\zeta/E_\nu(T)$
is an invariant probability measure under the dynamics of 
$\mathbf{\Phi}$, therefore, by uniqueness,
$\bar{\zeta}=\pi$, as required.
\qed



\bibliographystyle{plain}

\def\cprime{$'$}


\end{document}